# Focused-ion-beam milling
## based nanostencil mask fabrication for spin transfer torque studies

B. Özyilmaz[a], G. Richter, N. Müsgens, M. Fraune, M. Hawraneck, B. Beschoten[b], and G. Güntherodt

Physikalisches Institut IIA, RWTH Aachen, and Virtual Institute for Spinelectronics (VISel), 52056 Aachen, Germany

M. Bückins and J. Mayer

Gemeinschaftslabor für Elektronenmikroskopie, RWTH Aachen, 52056 Aachen, Germany

Focused-ion-beam milling is used to fabricate nanostencil masks suitable for the fabrication of magnetic nanostructures relevant for spin transfer torque studies. Nanostencil masks are used to define the device dimensions prior to the growth of the thin film stack. They consist of a wet etch resistant top layer and an insulator on top of a pre-patterned bottom electrode. The insulator supports a hard mask and gives rise to an undercut by its selective etching. The approach is demonstrated by fabricating current perpendicular to the plane Co/Cu/Co nanopillar junctions, which exhibit current-induced magnetization dynamics.

---

[a] Present address: Department of Physics, Columbia University, New York, New York 10027, USA

[b] Electronic mail: bernd.beschoten@physik.rwth-aachen.de

In recent years focused-ion-beam (FIB) milling has been increasingly used to fabricate magnetic nanostructures. The versatility of resist free FIB patterning allows the fabrication of a wide variety of nanostructures in the sub-100 nm size range[1, 2]. Most approaches utilize FIB milling as a subtractive method. The desired device structure is cut out of a ferromagnetic thin film thus producing mainly planar devices such as nano-constrictions in a magnetic film or arrays of ferromagnetic nanometer size elements[3]. A general concern in the usage of FIB is the influence of focused-ion-beam induced change on device performance[4]. While irradiation effects may be reduced significantly by optimizing the process parameters, it is material and device geometry dependent. Thus, FIB patterning requires additional steps for any changes in either material combination or device geometry, making the subtractive approach in many cases too slow for quick material screening and the exploration of critical system parameters. Furthermore, while the impact of the irradiation on device performance may be minimized, its influence on the fundamental properties in particular when probed by transport measurements is poorly understood. It is well known, that FIB milling leads to ion implantation, introduces magnetic pinning defects and leads to nanometer-scale displacements[5, 6]. Thus, these effects may mask intrinsic properties of the system to be studied. For example, spin transfer torque[7, 8] driven magnetization reversal[9], magnetization precession[10] or domain wall motion[11] depend strongly on both the local magnetization and local conductivities.

Here we demonstrate an alternative approach to FIB-based fabrication of magnetic nanostructures suitable for utilizing and probing intrinsic properties of magnetic nanostructures in both the in-plane and out-of-plane configuration. Rather than defining the dimensions of the nanoscale feature by removing material, in our novel approach the dimensions of the nanomagnets are defined prior to their deposition. The latter is achieved by means of FIB fabricated nanostencils. With a sufficient undercut in the nanostencil template our approach is particularly well suited for the fabrication of layered structures such as spin valves in the current perpendicular to the plane geometry (CPP). Note that in the case of the pillar junction fabrication the nanostencil mask approach has been successfully demonstrated via standard e-beam lithography[12]. Here we show that the FIB is equally well suited. Similar to Ref. [12, 13], our approach allows the fabrication of large arrays of templates. One can now quickly modify and optimize both, material combinations and growth conditions. This approach is not limited to ferromagnetic structures, but may also be of interest to other nanometer size layered structures where special attention must be paid to the FIB-induced changes in material properties and material composition.

From a process flow point of view, the approach described here differs from Ref. [12, 13] such that the bottom electrode is fabricated prior to the fabrication of the nanostencil mask. This has two advantages. First, the turn around time after ferromagnetic thin film growth is reduced considerably. Second, the nanostencil mask approach can now also be used to fabricate spin transfer torque devices which operate in the current in plane (CIP) geometry [11]. For this geometry it is sufficient to replace the (single) bottom electrode of a pillar junction by two or more electrodes of desired separation, such that the latter can act as current and voltage probes. FIB etching is a direct patterning technique. Thus, in the absence of pattern transfer issues, feature sizes can be reduced in principle into the 10 nm range and below. Since many intriguing effects in spin transfer torque rely on very high current densities[14], this may help to study the latter not only in the high current bias regime but also in the ballistic regime. Here we illustrate our method by fabricating a traditional bilayer pillar junction, the model system for most spin transfer torque studies.

Our process starts with the fabrication of the nanostencil mask consisting of a thin film stack of three layers. From top to bottom these are the hard mask layer, a thin insulator and an inert bottom layer, which acts as the bottom electrode. The choice of material combination has to be such that the insulator can be etched selectively without influencing either mask material or the bottom electrode(s). One choice of material combination for the fabrication of spin transfer torque devices is a Pt|SiO$_2$|Pt trilayer. Here Pt acts as both the mask material and the bottom electrode material. SiO$_2$ is the insulating layer, which is etched isotropically via a HF dip. The relevant process steps are summarized in Fig. 1. First we fabricate the bottom electrode(s) by means of optical lithography and subsequent Ar ion milling. The process continues with the sputter deposition of the insulator and the hard mask layer. Next FIB[c] is used to open up an aperture in the top Pt layer (Fig. 1(a)). This process step defines the critical dimensions (10 - 100 nm) of the junction and gives access to the SiO$_2$ layer for the subsequent HF dip. In order to avoid damage to the bottom electrode, the FIB milling step is stopped close to the first Pt|SiO$_2$ interface. This is easily achieved, since Pt etches approximately four times faster than SiO$_2$; the latter acts effectively as an etch stop for the FIB-ion milling step. Subsequently, isotropic wet etch of the SiO$_2$ with buffered HF is used to generate an undercut in the SiO$_2$ (Fig. 1(b)). This is the crucial step in the nanostencil mask fabrication, since it provides the necessary magnetic and electric separation to the ferromagnetic layer within the top electrode. Next the nanostencil is transferred into the HV sputter system and the desired thin film stack is deposited (Fig.1(c)). The subsequent process step depends on whether the final device is a CIP or CPP structure. In the case of the latter, the undercut is filled up with a thick contact metal layer. In the case of the CIP structure, the undercut can be kept open after the growth of the thin film stack to have access for scanning probe measurements. As a last step, the sample then goes through a single conventional optical lithography step with a subsequent Ar-ion milling defining the top electrode (Fig. 1(d)). The Ar-ion milling step gives also access to the bottom electrode[d].

---

[c] The FEI Strata 205 with a Ga+ liquid metal ion source has been operated at 30 keV with a beam current of 1pA.

A good indication of the presence of an undercut can be obtained from scanning electron microscopy (SEM) top views of the aperture as a function of etch time (Figs. 2(a-d)). In Fig. 2(a) a 140 nm × 140 nm opening is shown immediately after the FIB milling step. In Figs. 2(b-d) the etch times are 60 s, 180 s and 240 s, respectively. In Fig. 2(e) we show an array of FIB patterned undercut openings with a subsequent HF dip of 300s, for which the lateral dimensions range from 30 nm × 50 nm up to 100 nm × 350 nm.

As an example we discuss transport data obtained with a 70 nm × 140 nm pillar junction in more detail. Transport measurements with a CIP device will be discussed elsewhere[15]. The stack sequence for the pillar junction from bottom to top is |10 nm

---

[d] In both configurations the insulator is thin enough such that the bottom electrodes can be accessed by punching through it via wire bonding.

Pt|170 nm Cu|3nm Co|11 nm Cu|13nm Co|17 nm Cu|. Transport measurements were conducted at room temperature. The differential resistance $dV/dI$ was measured by lock-in technique with a 100 μA modulation current at $f = 1132$ Hz added to a DC bias current. Positive current is defined such that the electrons flow from the thin to the thick ferromagnetic layer.

In Fig. 3(a) we show a characteristic low DC bias magnetoresistance measurement of our asymmetric pillar junction at room temperature and with an in-plane field applied along the easy axis. A stack resistance of ∼ 1 Ω is typical for FIB fabricated pillar junctions. The device exhibits a clear transition between a low resistance state and a high resistance state corresponding to the parallel and anti-parallel configuration of the ferromagnetic layers in the junction. The giant magnetoresistance (GMR) value is ΔR/R ≈ 0.4 %. Next we discuss current sweeps at fixed magnetic fields. Characteristic current sweep traces for selected field values are shown in Fig. 3(b) and demonstrate current induced magnetization reversal. Comparing Fig. 3(a) with Fig. 3(b) we see that the current induced change in junction resistance (ΔR/R ≈ 0.4 %) is similar to the field induced change in junction resistance. The threshold currents for current induced magnetization reversal increase with increasing applied field. This is summarized in Fig. 3(c) where we have plotted the difference in differential resistance for current sweep up and current sweep down on a gray scale as a function of both the applied field and the current bias. Here the current is swept from -20 mA to +20 mA and back to -20 mA, and the magnetic field is stepped from -600 Oe to + 750 Oe, with ΔH= 10 Oe.

In summary, an alternative approach to FIB patterning assisted preparation of magnetic nanostructures is demonstrated. It allows the fabrication of nanostencil mask templates for fast exploration of spin transfer torque effects in magnetic thin films in both the CIP and CPP geometry in the sub-100 nm size range, without the limitations and concerns usually associated with FIB milling. The approach is demonstrated by fabricating CPP pillar junctions, which have sufficiently low contact resistance to study spin transfer torque induced magnetization dynamics. The devices show a GMR value of 0.4 % at room temperature and exhibit current-induced magnetization reversal.

One of us (B. Ö.) gratefully acknowledges useful discussions with J. Z. Sun and A.D. Kent. This work was supported by DFG/SPP 1133 and by HGF.

**Fig. 1: Process flow: (a) First the bottom electrodes (Pt) are patterned with optical lithography. Next an insulator (SiO2) and a hard mask (Pt) are sputter deposited. FIB is used to open up the hard mask. (b) A wet etch generates an undercut and gives access to the bottom electrode. (c) The desired magnetic multilayer (MML) thin film stack is deposited next. (d) Optical lithography is used to define the top electrode and allows access to the bottom electrode via wire bonding.**

**Fig. 2: (a-d) SEM top of view of Pt opening after FIB milling, from left to right etch times are 0 s, 60 s, 180 s and 240 s. (e) Array of undercut openings after 240 s HF dip. Smallest opening is 30 nm × 50 nm in size.**

**Fig. 3: (a) Differential resistance as a function of field sweep at zero dc bias and (b) current sweep at selected field values. (c) Gray scale plot of the difference in differential resistance *dV/dI* for current sweep up and current sweep down as a function of both current bias and applied field. Dashed arrow indicates field step direction.**

[1] J. Gierak *et al*., Microelectronic Engineering, **78-79**, p. 266-278 (2005).

[2] A. A. Tseng, Small, No.**10**, 924-939 (2005).

[3] G. Xiong *et al*., Appl. Phys. Lett. **79**, 3461 (2001).

[4] M. Kläui *et al*., Microelectronic Engineering **73-74**, 785-789 (2004).

[5] P. Warin *et al*., J. Appl. Phys. **90**, 3850 (2001).

[6] R. Hyndman *et al*., J. Appl. Phys. **90**, 384 (2001).

[7] J. Slonczewski, J. Magn. Magn. Mater. **159**, L1(1996).

[8] L. Berger, Phys. Rev. B **54**, 9353 (1996).

[9] J. Z. Sun, J. Magn. Magn. Mater. **202**, 157 (1999); J.-E. Wegrowe *et al*., Europhys. Lett. **45**, 626 (1999); J. A. Katine *et al*., Phys. Rev. Lett. **84**, 3149 (2000); J. Grollier *et al*., Appl. Phys. Lett. **78**, 3663 (2001); S. Urazhdin *et al*., Phys. Rev. Lett. **91**, 146803 (2003); B. Özyilmaz *et al*., Phys. Rev. Lett. **91**, 067203 (2003).

[10] M. Tsoi *et al*., Phys. Rev. Lett. **80**, 4281 (1998); Y. Ji, C. L. Chien and M. D. Stiles, Phys. Rev. Lett. **90**, 106601 (2003); S. I. Kiselev *et al*., Nature, **425**, 380 (2003); B. Özyilmaz *et al*., Phys. Rev. Lett. **93**, 176604 (2004); W. H. Rippard *et al*., Phys. Rev. Lett. **92**, 027201 (2004); I. N. Krivorotov *et al*., Science **307**, 228 (2005).

[11] M. Tsoi *et al.*, Appl. Phys. Lett. **83**, 2617–2619 (2003); M. Kläui *et al*., Phys. Rev. Lett. **94**, 106601 (2005); G. S. D. Beach *et al*. Nature Materials **4**, 741 (2005).

[12] J.Z. Sun e*t al.*, Appl. Phys. Lett. **81**, 2202 (2002).

[13] J. Z. Sun *et al*., J. Appl. Phys. **93**, 6859 (2003).

[14] B. Özyilmaz *et al*., Phys. Rev. Lett. **93**, 176604 (2004); M. L. Polianski and P. W. Brouwer, Phys. Rev. Lett. **92**, 26602 (2004); M. D. Stiles, J. Xiao and A. Zangwill, Phys.

Rev. B **69**, 054408 (2004); A. Brataas, Y. Tserkovnyak and G. E. W. Bauer, cond-mat/**0501672** (2005); S. Adam, M. L. Polianski and P. W. Brouwer, cond-mat/**0508732** (2005).

[15] G. Richter *et al.,* to be published.

**Focused - ion - beam**

**Wet etch**

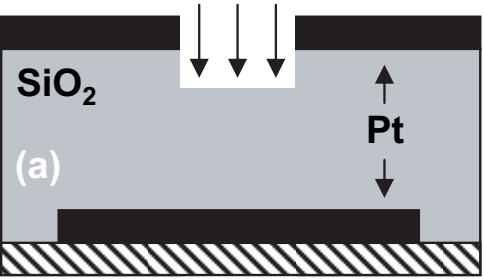

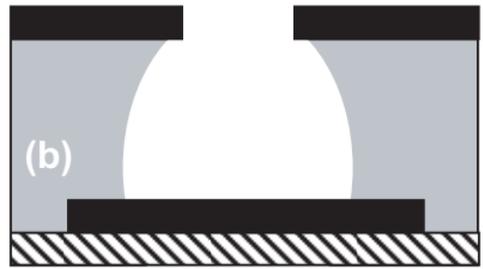

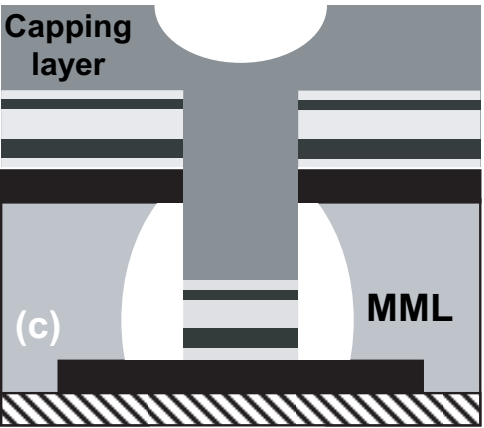

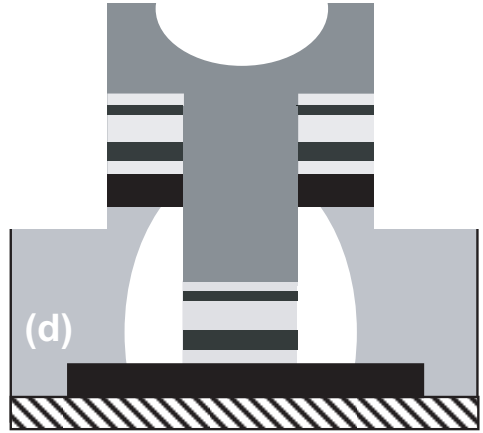

Figure 1: B. Özyilmaz et al.

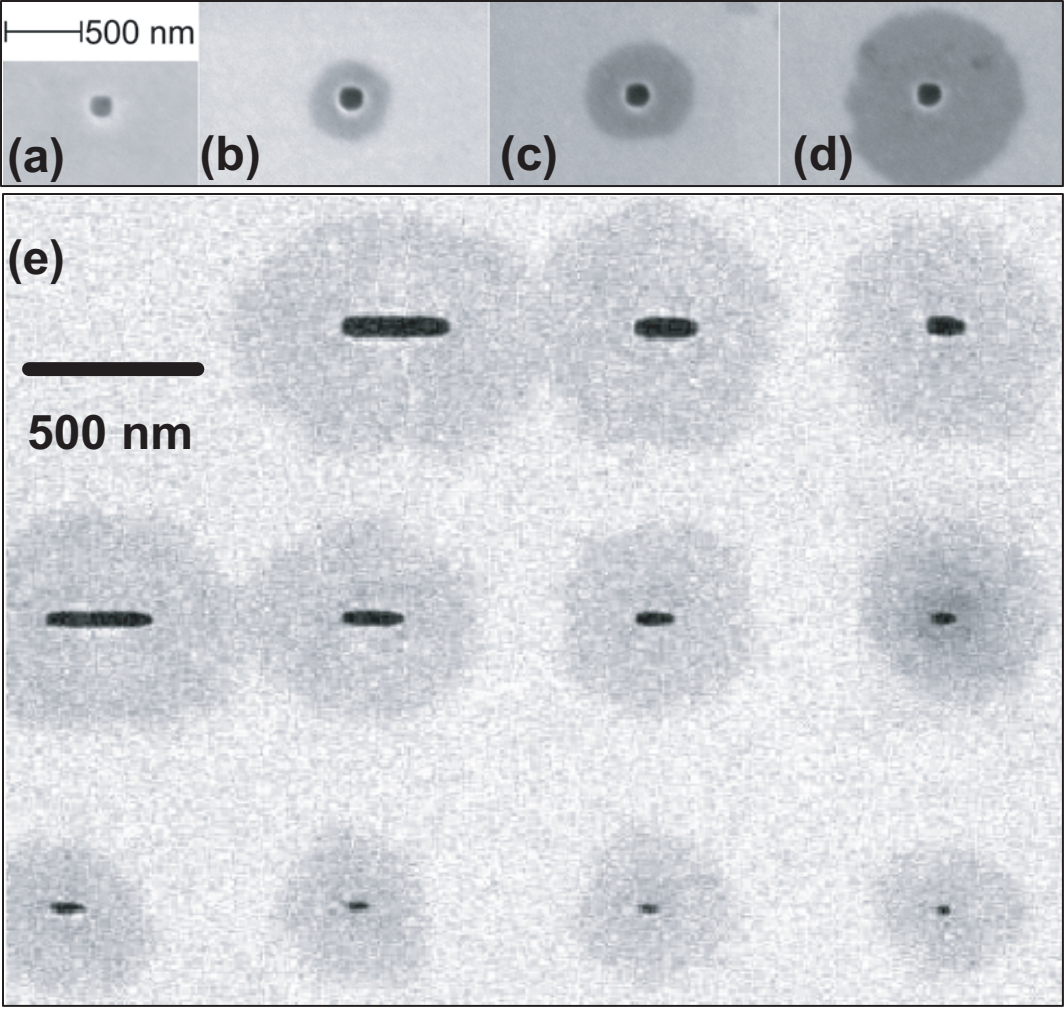

500 nm

(a)　(b)　(c)　(d)

(e)

500 nm

Figure 2: B. Özyilmaz et al.

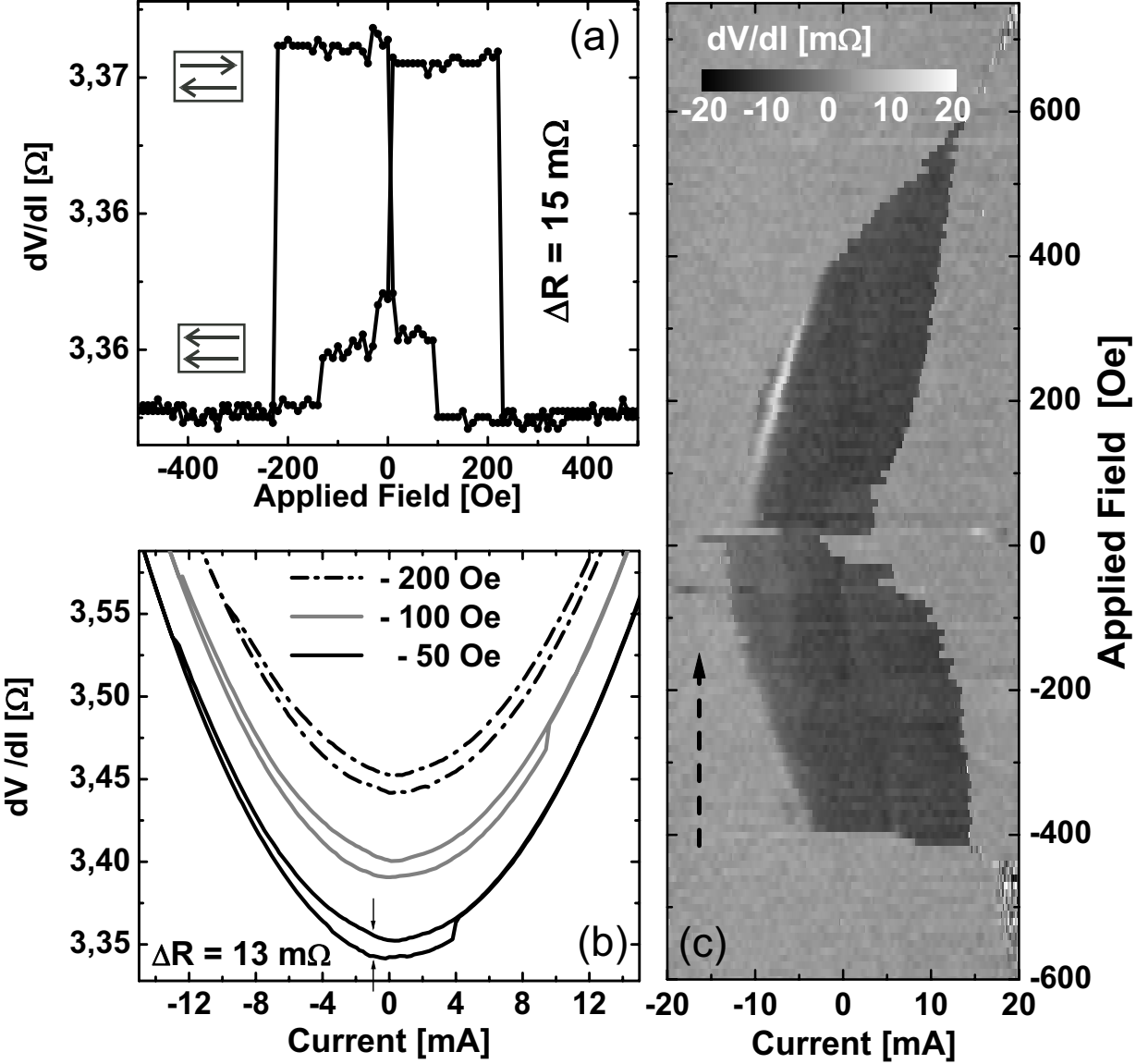

Figure 3: B. Özyilmaz et al.